\documentclass[12pt,a4paper]{article}

\usepackage[colorlinks=true, urlcolor=blue, linkcolor=blue, citecolor=blue]{hyperref}
\usepackage{graphicx}

\newcommand{\beq}{\begin{equation}}
\newcommand{\eeq}{\end{equation}}
\newcommand{\beqa}{\begin{eqnarray}}
\newcommand{\eeqa}{\end{eqnarray}}
\newcommand{\beqar}{\begin{eqnarray*}}
\newcommand{\eeqar}{\end{eqnarray*}}

\begin{document}
$\;$
\vspace{10pt}
\begin{center}

{\textbf{\Large Production and propagation of heavy hadrons}}

{\textbf{\Large in air-shower simulators}}

\vspace{40pt}

C.~A.~Garc\'\i a Canal$^1$, J.~I.~Illana$^2$, M.~Masip$^2$, S.~J.~Sciutto$^1$
\vspace{12pt}

\textit{$^1$IFLP/CONICET and Departamento de F{\'\i}sica}\\ 
\textit{Universidad Nacional de La Plata, C.C.67, 1900, 
La Plata, Argentina}\\
\vspace{10pt}
\textit{$^2$CAFPE and Departamento de F{\'\i}sica Te\'orica y del
Cosmos}\\ 
\textit{Universidad de Granada, E-18071, Granada, Spain}\\
\vspace{16pt}
\texttt{cgarciacanal@fisica.unlp.edu.ar, jillana@ugr.es, masip@ugr.es, 
sciutto@fisica.unlp.edu.ar}
\end{center}

\vspace{40pt}

\date{\today}

\begin{abstract}

Very energetic charm and bottom hadrons may be produced 
in the upper atmosphere when a 
primary cosmic ray or the
leading hadron in an extensive air shower collide with a
nucleon. 
At $E\approx 10^8$~GeV their decay length becomes of order 
10~km, implying that they tend to interact in the
air instead of decaying. Since 
the inelasticity in these collisions
is much smaller than the one in proton and pion collisions, 
there could be rare events where a 
heavy-hadron component 
transports a significant amount of energy 
deep into the atmosphere. We have developed a module for 
the detailed simulation of these processes and have included it
in a new 
version of the air shower simulator AIRES.
We study the frequency, the energy distribution and
the depth of charm and bottom production and decay in
the atmosphere. 
As an illustration, we consider the 
production and decay of tau leptons (from $D_s$ decays) 
and the lepton flux at PeV energies from a 30~EeV proton 
primary.
The proper inclusion of charm and bottom hadrons 
in AIRES opens the
possibility to search for air-shower observables 
that are sensitive to heavy quark effects.
\end{abstract}


\newpage

\section{Introduction}

Air-shower simulations are an essential tool in 
cosmic-ray physics \cite{AIRES,Knapp:2002vs}. Primary particles 
reach the Earth with energies of up to
$10^{11}$~GeV, in particular, AUGER is exposed to around 15,000
events of energy above $10^{10}$~GeV ($10\%$ of them hybrid)
per year \cite{Settimo:2012zz}. Such energies are well 
above the ones explored at colliders,
and the simulation of these events requires then an 
extrapolation of the {\it known} physics that could
be affected by several factors. On one hand, there could
be new particles or interactions not accessible 
at lower energies. In this sense, cosmic rays may offer 
opportunities in the search for strong TeV 
gravity \cite{Emparan:2001kf}, new
neutrino interactions \cite{Anchordoqui:2002vb},
or long-lived massive particles \cite{Albuquerque:2003mi}.
On the other hand, cosmic-ray energies may imply a regime
where the properties of standard
particles can be substantially {\it different}.
Consider, in particular, the hadrons containing a charm
or a bottom quark, whose properties 
are well known from collider experiments.
The lightest mode with a given quark content will 
always decay through a weak interaction, implying
a relatively long lifetime ($c\tau=0.1$--$0.5$ mm).
Although at the Tevatron or the LHC very energetic
heavy hadrons may define events with 
observable displaced vertices, such  hadrons 
will never reach the calorimeters there. In contrast,
when produced with energies above $10^8$~GeV 
inside an extensive air shower (EAS) 
their decay length becomes larger than 10 km,
and they tend to collide in the atmosphere
instead of decaying. 
The collisions with matter of these long-lived heavy
hadrons would introduce physics unseen at
colliders. This physics will certainly occur in 
extensive air showers, and its inclusion in the
simulation may be necessary to explain 
rare effects or just as a standard background in the search for
genuine exotics.

The heavy quark inside an
ultrarelativistic $D$ or $B$ meson
carries most of the hadron energy. If the meson collides with
an air nucleus and {\it breaks} into several pieces, it is
obvious that the piece carrying the heavy quark will 
take most of the energy after the collision. Therefore,
at $E> 10^8$~GeV 
these particles become long lived and
much more penetrating than a proton or a pion: 
a simulation seems necessary to establish whether 
heavy mesons are effective in taking a significant fraction
of this energy deep into the atmosphere.

In addition, at energies above 100~GeV pions and kaons 
become less effective producing atmospheric muons and
neutrinos (they tend to collide with air nuclei  
instead of decaying), and the spectral index in the lepton
flux that they yield is reduced in one unit \cite{Lipari:1993hd}. 
The prompt decay of charmed hadrons has
been extensively studied \cite{Costa:2000jw}
as the dominant source of leptons at PeV
energies (see
also \cite{Illana:2010gh}). The non-prompt charm contribution  
(from charm decaying after one or
several collisions in the air), however, 
may be not negligible, specially at
higher energies, and its inclusion in the simulation 
requires an estimate of propagation effects.

In this article we report on the inclusion of heavy-quark production 
and propagation in the air-shower simulator AIRES \cite{AIRES}.
AIRES provides full space-time shower development
in a realistic environment, taking
into account the atmospheric density profile, 
the Earth's curvature, and the geomagnetic field.
The new version of the simulator used in this work
recognizes and propagates 
photons, electrons, positrons, muons, neutrinos,
pions, kaons, eta mesons, lambda baryons, nucleons, antinucleons, 
nuclei up to $Z=36$, as well as 
$D$ and $B$ mesons, $\Lambda_c$ baryons, and   
tau leptons (which may appear in $D_s$ decays). AIRES is 
able to process complex decay schemes with a large 
number of branches, as it is the case for heavy hadrons. 
Nucleus-nucleus, hadron-nucleus, and photon-nucleus
inelastic collisions with significant cross-sections are taken into
account via calls to external hadronic packages.

For the present work we have developed a hadronic interaction 
preprocessor (HQIP, for Heavy Quark Interaction Preprocessor)
that simulates collisions including heavy hadrons. 
The algorithms used in HQIP are based on a perturbative QCD 
framework that is described in some 
detail in section \ref{sect:charmbot}. 
When processing high energy hadronic collisions, HQIP complements 
the usual hadronic 
packages (SIBYLL~\cite{Fletcher:1994bd} or QGSJET~\cite{Ostapchenko:2006}).
If HQIP is invoked the charm production option 
in the external hadronic package is set to disabled.
The production of heavy quarks in 
nucleus-nucleus and photon-nucleus 
collisions has not been implemented yet.

The impact of charm production in air showers has been previously
discussed in \cite{dietercharm}. That work, however, 
does not include bottom quark production nor the propagation 
of heavy hadrons in the atmosphere ({\it i.e.}, their possible 
collisions with air nuclei).

This article is organized as follows. 
In section \ref{sect:charmbot} we discuss the
production cross sections and the inelasticities 
in heavy meson collisions with air nuclei that we have used.
In section \ref{sect:showerdevel} we illustrate the performance
of AIRES with a 
study of the energy and depth distributions 
of heavy hadron production and decay for a vertical
proton primary of fixed $30\mbox{ EeV}=3\times 10^{10}$~GeV
energy. Finally, we discuss the spectrum of muons 
and the production of very
energetic tau leptons from heavy quark decays 
for the same proton primary.

\section{Charm and bottom production and propagation}
\label{sect:charmbot}

Heavy-quark production by cosmic rays has been considered
by a number of groups (see 
\cite{Berghaus:2007hp} for a review). At ultrahigh
energies the usual
calculation of the partonic process using perturbative QCD
is very dependent on the small-$x$ gluon distribution, giving
results that can vary by more than an order of magnitude 
\cite{Costa:2000jw}. Instead, we will base
our cross sections on the color dipole picture 
\cite{Nikolaev:1990yw} described in detail
in \cite{Goncalves:2006ch,Enberg:2008te}, which 
incorporates in a simple way saturation and 
nuclear effects and yields much smaller uncertainties. 
Within this picture
a gluon carrying a fraction $x_1$ of the projectile energy
fluctuates into a $Q\bar Q$ dipole that interacts coherently
with the gluon field in the target (the $x_2\ll 1$ gluons 
form a Color Glass Condensate) and evolves into hadrons. 
\begin{figure}[!t]
\begin{center}
\begin{tabular}{c}
\includegraphics[width=0.48\linewidth]{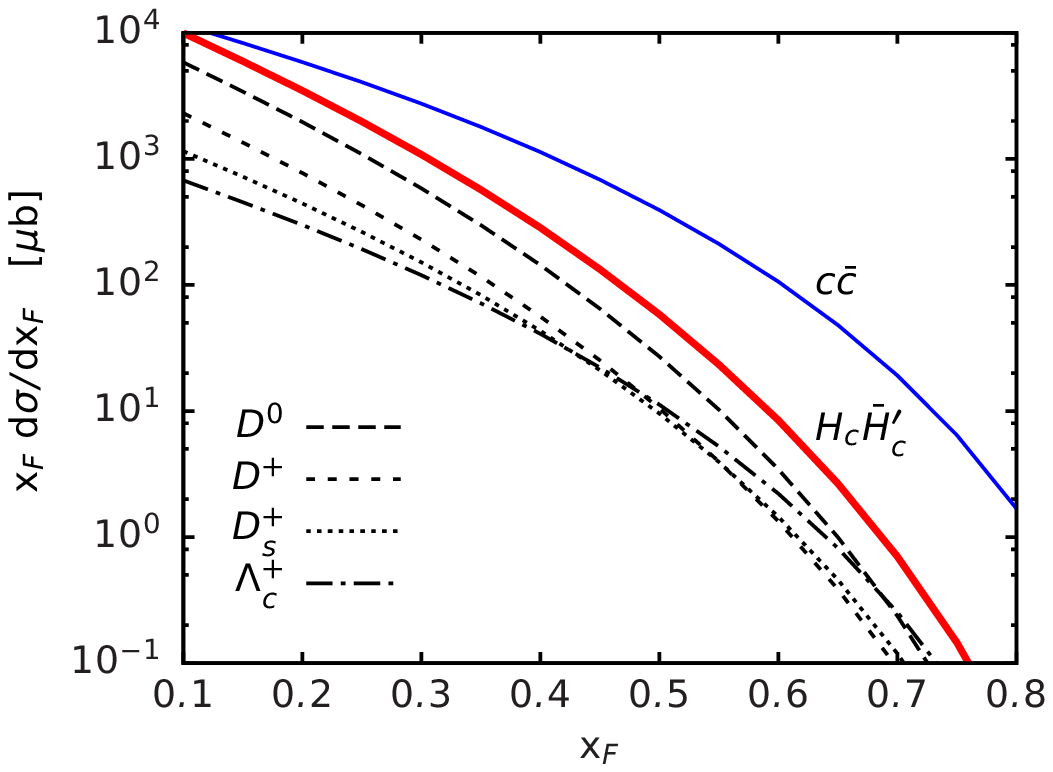}
\includegraphics[width=0.48\linewidth]{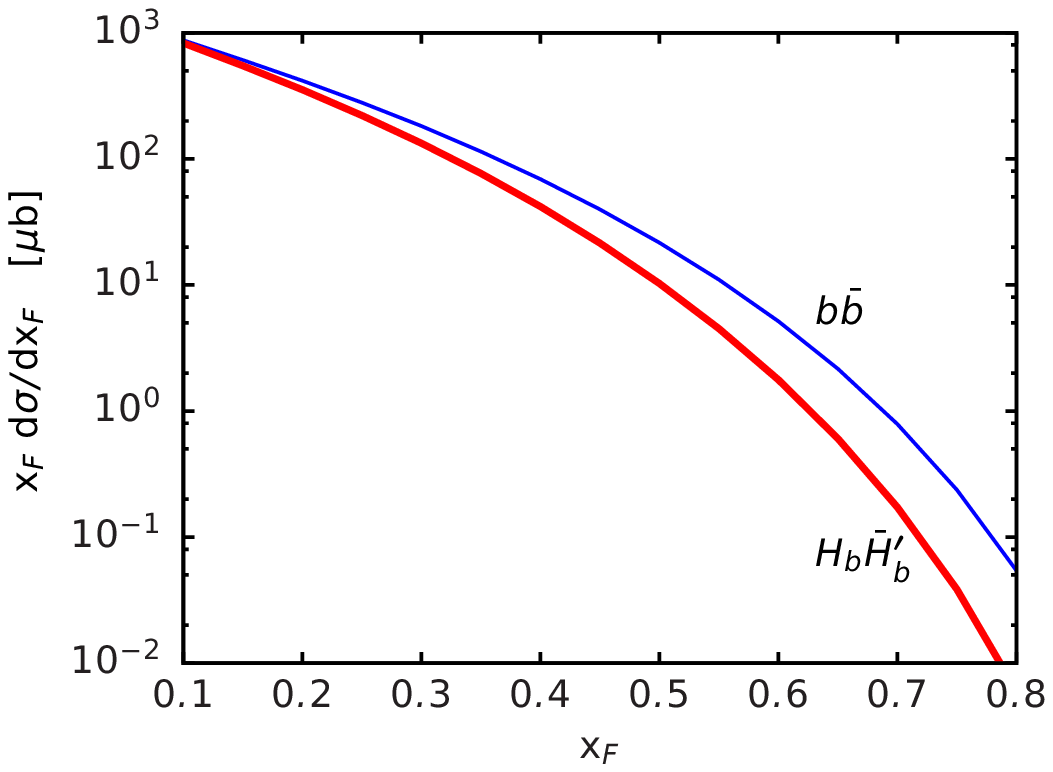} 
\end{tabular}
\end{center}
\caption{Differential cross section for $c\bar c$ (left)
and $b\bar b$ (right) production in $p$--air collisions for
an incident energy of $10^{10}$~GeV. The upper line gives the
fraction $x_F$ of energy taken by the $Q\bar Q$ pair, whereas 
the lower (thick) line gives the energy taken by the two
hadrons after fragmentation. In dashes we plot the cross section
for the production of each charmed species.
\label{fig1}}
\end{figure}
GM \cite{Goncalves:2006ch} and ERS \cite{Enberg:2008te} provide 
the inclusive differential cross section 
${\rm d} \sigma^{Q\bar Q}/{\rm d} x_F$ in proton--air collisions, 
where the Feynman variable $x_F=x_1-x_2\approx x_1$
gives the fraction of incident 
energy taken by the heavy-quark pair. Their results agree 
within the $50\%$ uncertainty that is expected from the choice 
of scales, PDFs, charm (or bottom) quark mass, and nuclear effects 
(see Fig.~3 in \cite{Enberg:2008te}). 
The new version of AIRES uses the simple parametrization  
in \cite{Goncalves:2006ch}.
We plot in Fig.~1 (upper lines) the differential cross
section for a $10^{10}$~GeV incident proton. 
The pair energy ($x_F E$) will be distributed
between the two heavy quarks. In particular, we assume
that the fraction 
$y$ of energy taken by each quark follows a flat distribution
between $y_{\rm min,max}=0.5(1\mp \sqrt{1-4\epsilon^2})$, with
$\epsilon=m_Q/m_{Q\bar Q}\approx 0.3$.

The heavy quarks will then fragment into hadrons. 
For the charm quark we have used the Kniehl and
Kramer parametrization \cite{Kniehl:2006mw}, including the fragmentation into
$D^0$, $D^+$ and $D_s^+$ mesons and $\Lambda_c^+$ baryons (plus 
the corresponding antiparticles for $\bar c$). 
In Fig.~\ref{fig1}--left we plot the fraction of energy taken
by the hadron pair ($H_Q\bar H_Q'$) together with the relative abundance 
of each species in $10^{10}$~GeV $p$--air collisions.
For the bottom quark (see \cite{Mele:1990yq}) 
we consider only fragmentation 
into $B^-$ or $\bar B^0$ mesons, with equal frequency and
a fragmentation function $D_b(z)=N z^{13.7}(1 - z)$, where
$z=E_H/E_Q$. In 
Fig.~\ref{fig1}--right we plot the fraction of energy taken by
the pair of $B$ mesons after the $b\bar b$ pair has been
produced.

\begin{figure}[!ht]
\begin{center}
\includegraphics[width=0.5\linewidth]{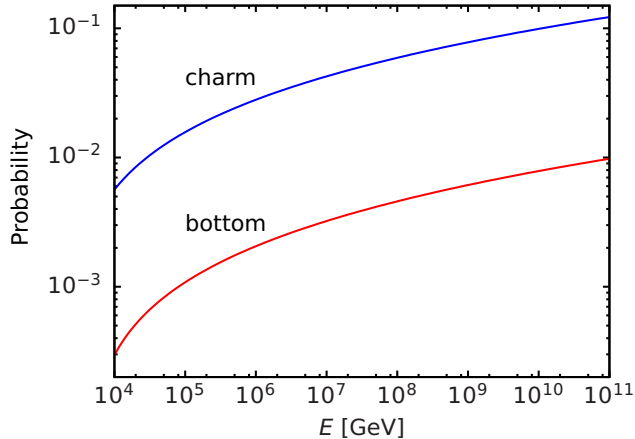} 
\end{center}
\caption{Probability to produce a charm or a bottom pair carrying 
more than $1\%$ of the proton energy in $p$--air collisions.
\label{fig2}}
\end{figure}
We have included in AIRES
the production of heavy-hadron pairs in nucleon and meson collisions 
with incident energy above $10$~TeV, and have considered only pairs 
carrying more than 1\%  of the projectile energy. 
These thresholds ensure the inclusion of all the effects 
that are relevant in air shower simulations and 
avoid complicating unnecesarily the procedure.
A pair of energy $E_{Q\bar Q}= x_F E$ may come from a parent hadron 
of energy $10 E_{Q\bar Q}$ in a collision of $x_F=0.1$, or from a 
$1000 E_{Q\bar Q}$ parent for $x_F=0.001$. Given the steep power law
observed in the cosmic ray flux, however, the contribution to 
the production of heavy quark pairs from 
small-$x_F$ collisions is not important.

Dividing by the total (inelastic) 
$p$--air cross section we obtain the probability
to produce pairs of charm or bottom hadrons 
with $x_F>0.01$. In 
Fig.~\ref{fig2}
we plot this probability for different incident energies
between $10^4$ and $10^{11}$~GeV. The same
production probability in the interactions 
of neutrons, charged pions and kaons with air has been assumed.

AIRES allows then the heavy hadrons either to decay or to 
collide with an
air nucleus. In the case of a collision, we have taken the
inelasticity and the interaction lengths for charm and bottom 
hadrons from \cite{Barcelo:2010xp} and 
\cite{Bueno:2011nt}, respectively. For example,
a $D$ meson after a $10^9$~GeV collision 
could keep around 55\% of the initial energy, 
whereas a $B$ meson
will typically exit with 80\% of the incident energy after
colliding with an air nucleus.
In contrast, the leading meson after a $10^9$~GeV pion collision 
would carry in average just
22\% of the energy.

\section{Heavy-quark production and evolution in air shower development}
\label{sect:showerdevel}

To study the heavy hadron production and evolution inside the
shower, we have generated 
10,000 vertical showers initiated by a 30~EeV proton primary
and have simulated them down to sea level (1000 g/cm$^2$)

\begin{figure}[!t]
\begin{center}
\begin{tabular}{c}
\includegraphics[width=0.48\linewidth]{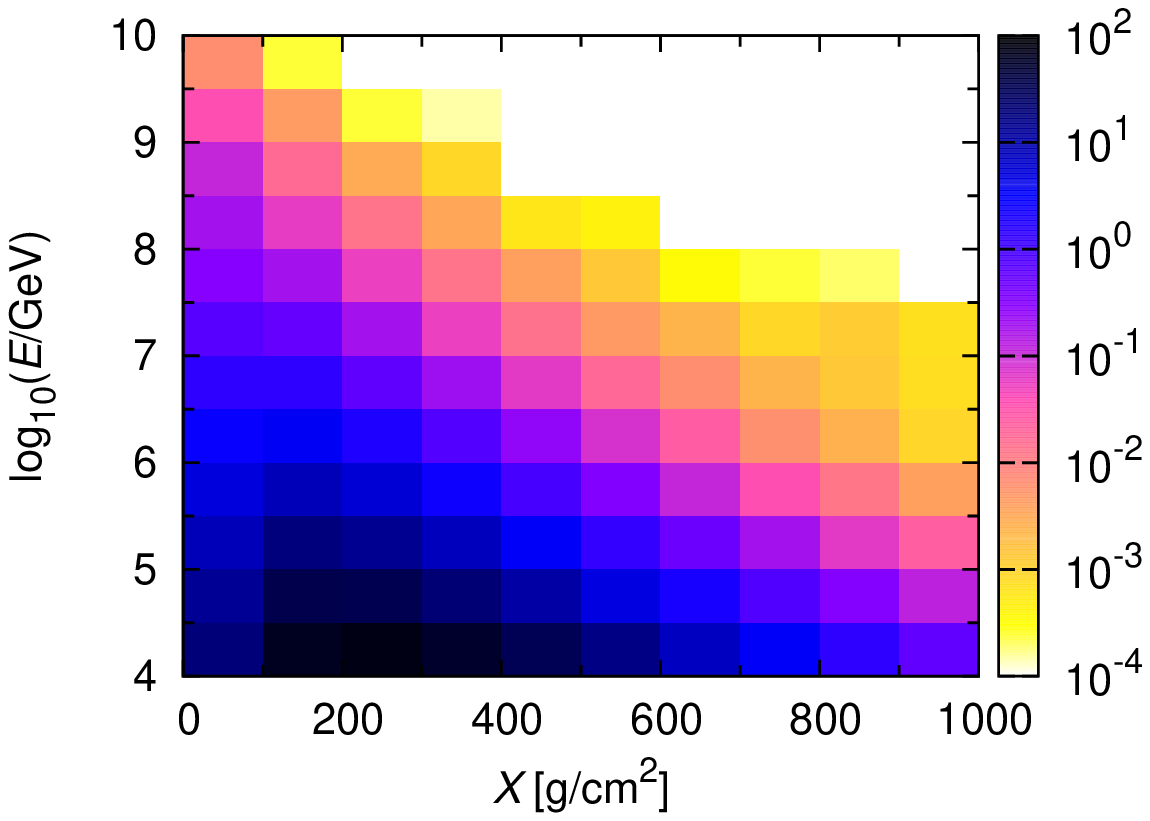}
\includegraphics[width=0.48\linewidth]{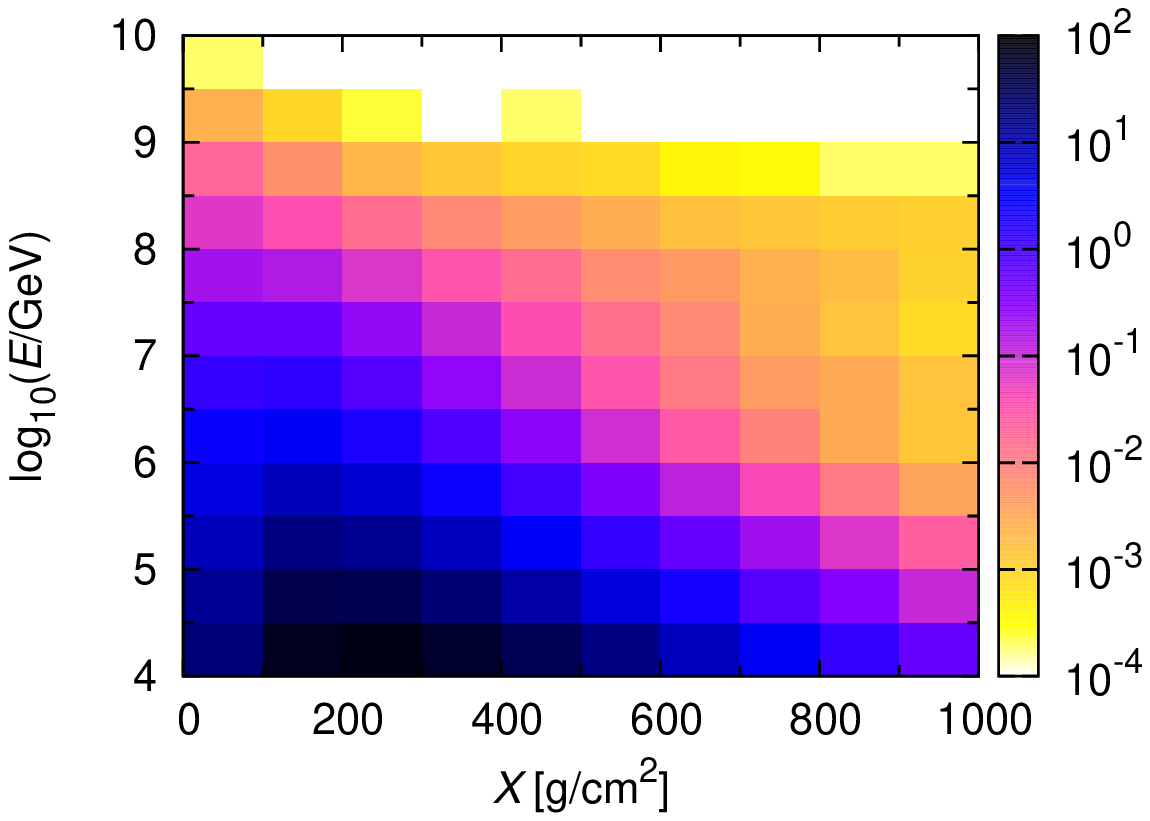} 
\end{tabular}
\end{center}
\caption{Average number of charm hadrons produced (left) and
decayed (right) per bin of 
energy and atmospheric depth.
\label{fig3}}
\end{figure}

In Fig.~\ref{fig3}--left we plot the average number of 
charmed hadrons produced per shower 
in bins of 100~g/cm$^2$ and half a decade of energy. For example,
from the plot it follows that around 0.19 charmed hadrons of 
$10^{7.5}$--$10^{8.0}$~GeV are produced at
atmospheric depths between $100$ and $200$~g/cm$^2$ per shower.
These hadrons are produced in secondary collisions 
and also through the decay of $B$ mesons (most of the high-energy 
charmed hadrons beyond 600~g/cm$^2$ come from $B$ decays).

Adding bins we find that an average 30~EeV 
shower contains
0.5 charmed hadrons of energy above $10^8$~GeV, and
6\% of them are produced beyond 200~g/cm$^2$.
The total energy transferred into these very energetic 
hadrons is $2.0\times 10^8$~GeV, i.e., 0.7\% of the energy of
the primary proton.
Concerning charmed-hadron decay, we plot in Fig.~\ref{fig3}--right its depth and
energy distributions. We find an average of 1.0  charmed hadrons
of energy above $10^5$~GeV 
decaying after 600~g/cm$^2$. They carry a 
total energy of $2.9\times 10^6$~GeV beyond that atmospheric depth.
\begin{figure}[!b]
\begin{center}
\begin{tabular}{c}
\includegraphics[width=0.48\linewidth]{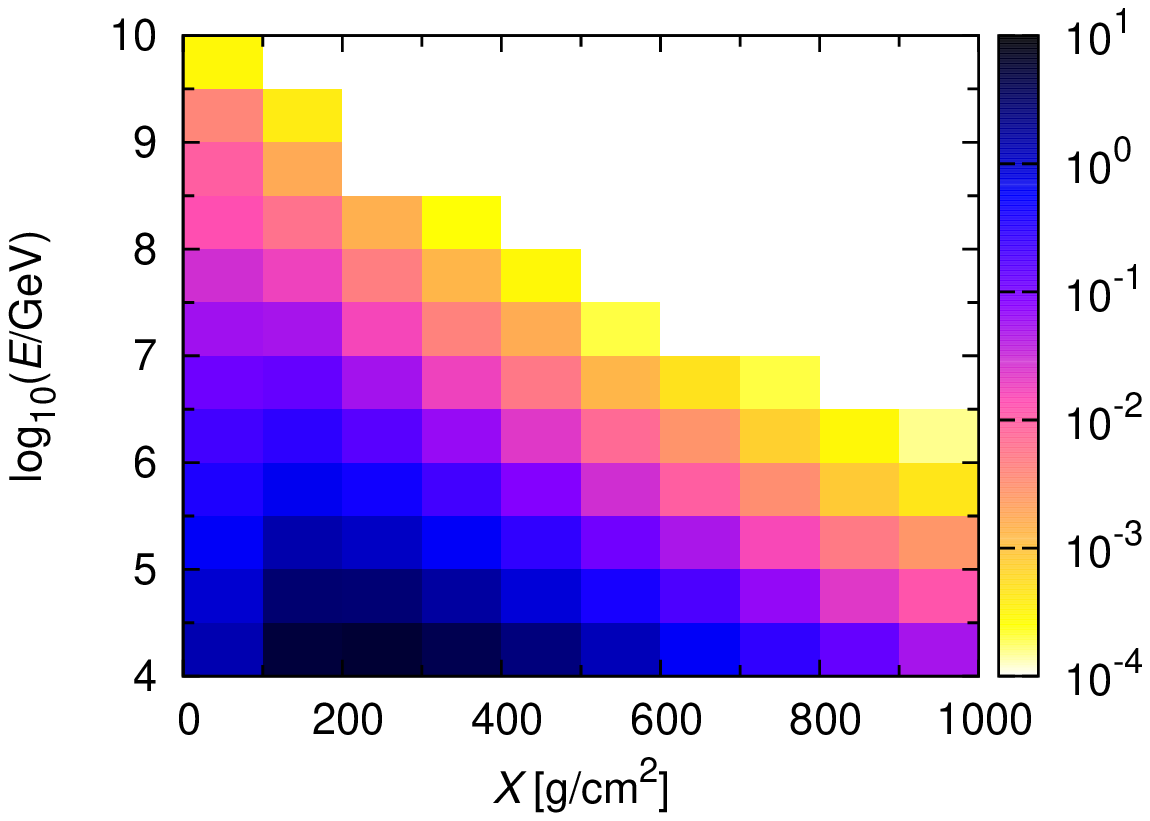}
\includegraphics[width=0.48\linewidth]{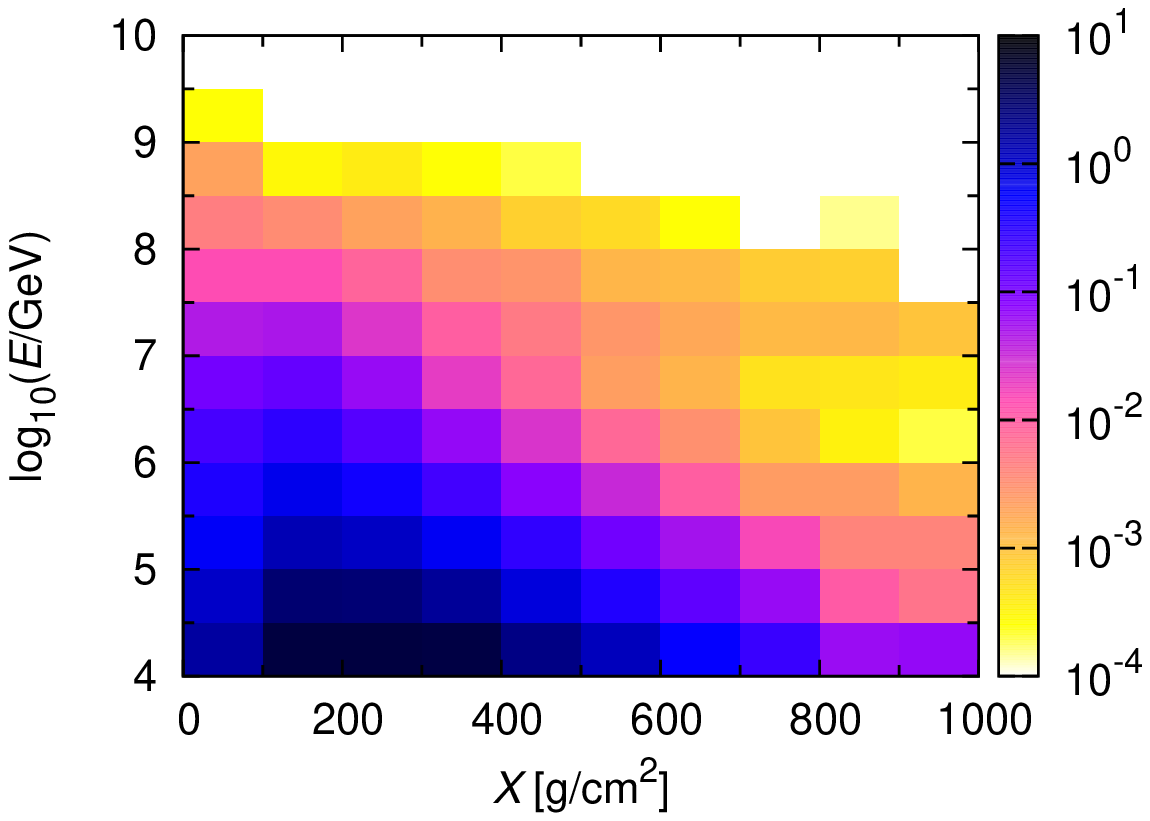} 
\end{tabular}
\end{center}
\caption{Average number of bottom hadrons produced (left) and
decayed (right) per bin of 
energy and atmospheric depth.
\label{fig4}}
\end{figure}

Bottom hadrons are more rare in these showers, but their
effect tends to be more {\it radical}. We plot the depth and 
energy distributions for their production 
(Fig.~4--left)
and their decay (Fig.~4--right) for 30~EeV vertical showers. 
We find around 38 $B$ mesons of energy 
above $10^8$~GeV per 1000 showers, 
with an average production depth of 
97~g/cm$^2$. This means that only 4\% of the 30~EeV showers 
include such an energetic bottom hadron. Given the higher 
elasticity in their collisions, these $B$ mesons reach deeper in
the atmosphere than charmed hadrons
before they decay. In particular, we find
around 16 hadrons per 1000 showers decaying at atmospheric depths beyond 
600~g/cm$^2$ with energies above $10^6$~GeV, and 1 hadron 
of energy above $10^7$~GeV reaching
the ground at 1000~g/cm$^2$.

\section{Tau leptons and muons}

In this section we present the frequency of 
tau lepton events and the average muon energy distribution from
our simulation of 30~EeV vertical proton showers.

\begin{figure}[!t]
\begin{center}
\begin{tabular}{c}
\includegraphics[width=0.48\linewidth]{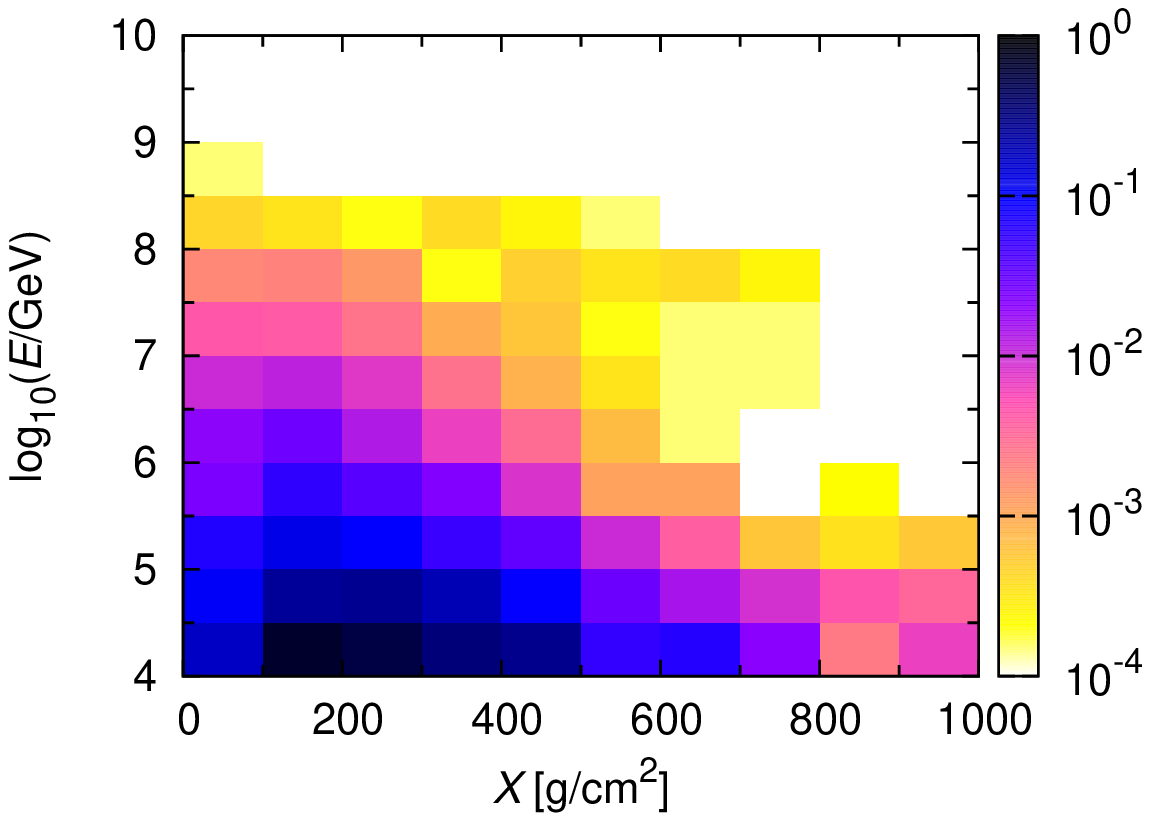}
\includegraphics[width=0.48\linewidth]{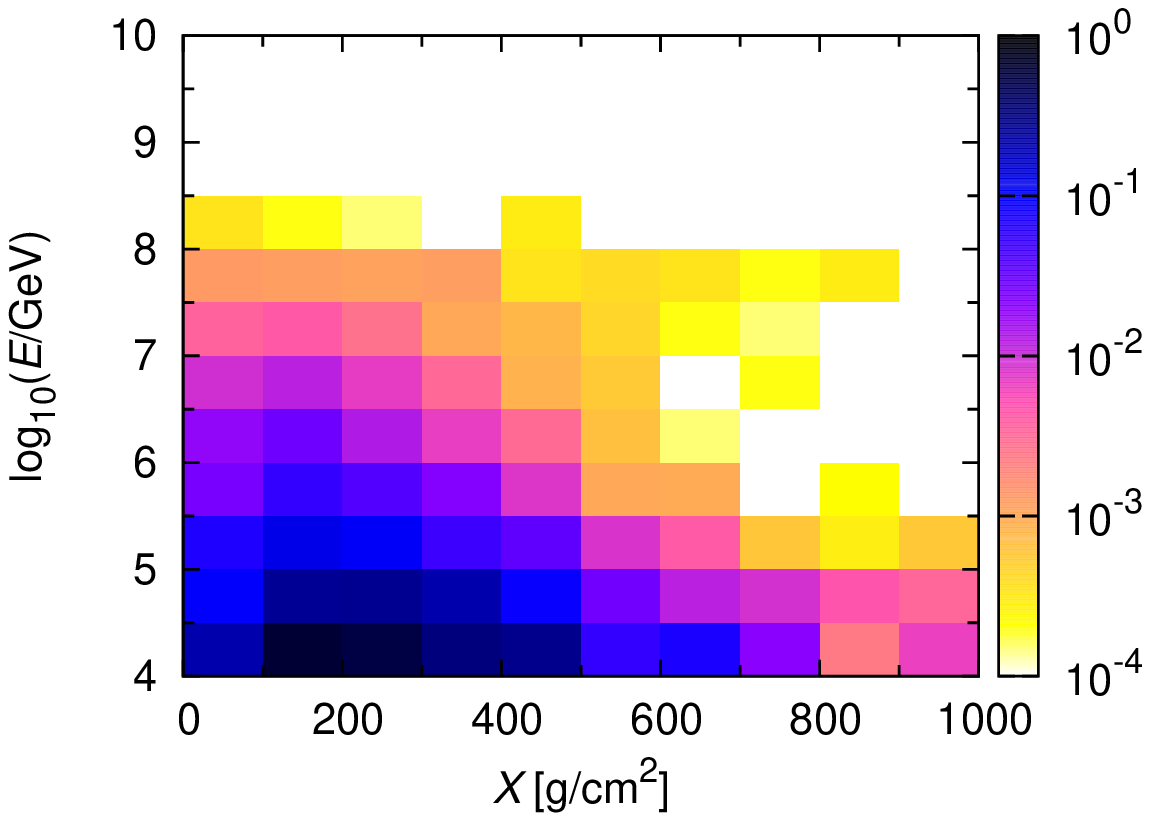} 
\end{tabular}
\end{center}
\caption{Average number of tau leptons produced (left) and
decayed (right) per bin of 
energy and atmospheric depth.
\label{fig5}}
\end{figure}
Tau leptons are mainly produced in
$D_s$ decays. They may introduce interesting effects because
their decay length ($c\tau=87$ $\mu$m) reaches 5 km at $10^8$~GeV.
In Fig.~\ref{fig5}--left (right) we plot the point and energy 
where they are produced (decay) inside the shower. 
We find 23 tau leptons of
energy above $10^7$~GeV per 1000 showers. They are produced at an average depth
of 270~g/cm$^2$ and decay at 320~g/cm$^2$, in particular, 
18\% of them decay after 600~g/cm$^2$. For example, 
in the energy bin between
$10^8$ and $10^{8.5}$~GeV we obtain 2.2 tau leptons per 1000 
vertical showers,
with 0.8 of them reaching the ground. Therefore, the frequency
of very energetic taus produced in EAS that decay near the ground 
is similar to the frequency of the 
analogous $B$-meson events.

Muons, on the other hand, are a key prediction in air-shower simulations. 
Although the presence of heavy hadrons will not introduce significant
differences in the {\em total} number of muons at the ground level, 
there are other observables that may be more
sensitive to these heavy hadrons.
In particular, one could expect two types of effects. 

\begin{figure}[!t]
\begin{center}
\includegraphics[width=0.5\linewidth]{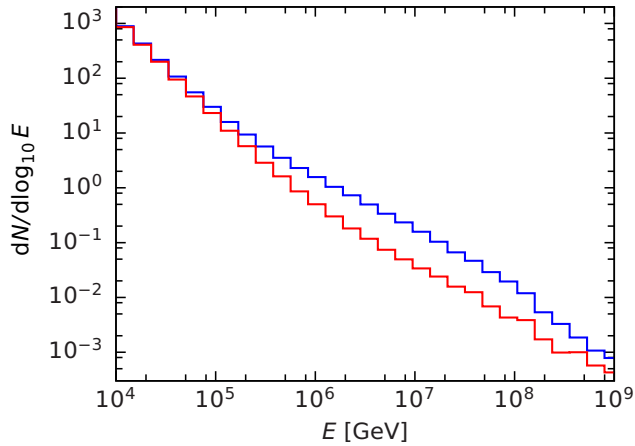} 
\end{center}
\caption{Average energy distribution of ground muons from 
a 30~EeV proton shower simulated with (upper histogram) and
without (lower histogram) the production, propagation and 
decay of heavy hadrons.
\label{fig6}}
\end{figure}

\begin{itemize}

\item
{\em Rare events with late energy deposition} 
from the decay of a heavy meson or a $\tau$ lepton. A  
$10^8$~GeV deposition relatively near 
the ground would produce muons and other charged particles 
that could change significantly  
the shower profile seen in the fluorescence telescopes
and/or the temporal distribution observed in the surface detectors.
The fraction of these rare events is approximately 0.5\%. 

\item
{\em Leptons of PeV energies}.
At very high energies pions tend to 
collide in the air instead of decaying thus becoming 
a less effective source of muons
\cite{Costa:2000jw,Illana:2010gh}.
This shows up clearly in 
Fig.~\ref{fig6}, where we plot
the energy distribution of muons reaching the
ground for a 30~EeV proton primary.
The distribution in
showers simulated using HQIP  
is almost 5 times larger than the one  
with no heavy quark production 
at PeV energies. These 
very energetic muons may be observable at neutrino 
telescopes \cite{Aguilar:2010kg},
and they could be used to estimate the correlated neutrino flux.

\end{itemize}

\section{Summary and discussion}

At energies above $10^8$~GeV hadrons containing a charm or a bottom quark
become long lived,  and it is then necessary to
implement their collisions with air nuclei in air shower simulations. 
Although the interactions with
matter of $D$ and $B$ mesons are not observable at colliders,
one expects that they are much more
elastic than pion or proton collisions. 

The main effect of these particles would derive from their ability
to transport energy deep into the atmosphere. Since the fraction of
events with a very energetic heavy hadron is small, one can
not expect significant differences in the features of the
average shower. In particular, we have studied common
observables like the lateral and the longitudinal distributions 
of charged particles, and in all cases we could not observe 
any relevant differences  due to heavy quark effects. 
Instead, one could look for anomalous events with
late energy deposition caused by their decay.

We have included 
both the production and the propagation of
heavy hadrons in a new version of AIRES.
To illustrate the performance of the code, we have simulated
10,000 vertical showers of fixed 30~EeV energy. 
We find around one $D$ meson of energy above $10^8$~GeV per 
2 showers, or just
one $B$ meson in this energy range per 26 showers. A few per mille 
of these air showers includes a $B$ meson of energy
above $10^7$~GeV hitting the ground at 1000~g/cm$^2$. The frequency
of very energetic tau leptons from $D_s$ decays reaching large
atmospheric depths is slightly higher. The uncertainty in all these
rates combines a $50\%$ variation in the production cross sections
with the uncertainty in the inelasticity of heavy hadron--nucleus
collisions, that is more difficult to estimate since 
such process can not be observed at colliders. 
Therefore, we think that a search for
possible signals associated to these rare events should be
considered even if the expected rates are small.

The results presented here point to different air shower
observables that may reflect the production of heavy quarks, 
but a definite determination of their observability would 
require the generation 
of air showers of different energies and from different inclinations.
The inclusion of heavy hadrons in AIRES
opens the possibility to optimize such a search.

\section*{Acknowledgments}
We would like to thank Javier Albacete, Antonio Bueno, and Alberto Gasc\'on for
useful discussions. 
The work  has been partially supported by 
ANPCyT and CONICET of Argentina, by 
MINECO of Spain (FPA2010-16802
and Consolider-Ingenio {\bf Multidark} CSD2009-00064)
and by Junta de Andaluc\'{\i}a (FQM 101, FQM 03048, FQM 6552).

\end{document}